\documentclass{article}
\usepackage{graphicx}
\def\be{\begin{equation}}
\def\ee{\end{equation}}
\def\bea{\begin{eqnarray}}
\def\eea{\end{eqnarray}}
\def\sh{\sinh}
\def\ch{\cosh}
\begin{document}
\title{
  {\bf Buda-Lund hydro model for ellipsoidally symmetric fireballs and
    the elliptic flow at RHIC}
}
\author{M. Csan\'ad$^{\, 1}$, T. Cs\"org\H{o}$^{\, 2}$ and B. L\"orstad$^{\, 3}$\\
    {\small $^1$ Department of Atomic Physics, ELTE University,}\\
  {\small H-1117 Budapest XI, P\'azm\'any P. 1/A, Hungary} \\
  {\small $^2$ MTA KFKI RMKI, H - 1525 Budapest 114, P.O.Box 49, Hungary}\\
  {\small $^3$ Department of Physics, University of Lund, S-22362 Lund, Sweden}
}
\maketitle
\begin{abstract}
The ellipsoidally symmetric extension of Buda-Lund hydrodynamic model
is shown here to yield a natural description of
the pseudorapidity dependence of the elliptic flow $v_2(\eta)$,
as determined recently by the PHOBOS experiment for Au+Au collisions
at $\sqrt{s_{NN}} = 130$ and $200$ GeV. With the same set
of parameters, the Buda-Lund model describes also
the transverse momentum dependence of $v_2$ of identified particles
at mid-rapidity. The results confirm the indication
for quark deconfinement in Au+Au collisions at RHIC, obtained from
a successful Buda-Lund hydro model fit to the
single particle spectra and two-particle correlation data, as
measured by the BRAHMS, PHOBOS, PHENIX and STAR collaborations.
\end{abstract}

\medskip
\medskip
\begin{flushright}
  {\it ``Nuclei, as heavy as bulls, \\
  through collision, generate\\
  new states of matter. " \\
  (T. D. Lee)}
\end{flushright}

\section{Introduction}
  Ultra-relativistic collisions of almost fully ionized Au atoms
  are observed in four major experiments at the RHIC accelerator
  at the highest currently available colliding energies of $\sqrt{s_{NN}}
  = 200$ GeV to create new forms of matter that existed before in Nature
  only a few microseconds after the Big Bang, the creation of our Universe.
  At lower bombarding energies at CERN SPS, collisions of Pb nuclei
  were studied in the $\sqrt{s_{NN}} = 17$ GeV energy domain, with a similar
  motivation. If experiments
  are performed near to the production threshold of a new state of matter,
  perhaps only the most violent and most central
  collisions are sufficient to generate
  a transition to a new state. However, if the energy is well
  above the production threshold,
  new states of matter may appear already in the mid-central or
  even more peripheral collisions.
  Hence the deviation from axial symmetry of the observed
  single particle spectra and two-particle correlation functions
   can be utilized to characterize the properties of such new states.

  The PHENIX, PHOBOS and STAR experiments at RHIC produced a
  wealth of information on the asymmetry of the particle spectra
  with respect to
  the reaction plane~\cite{PHENIX-v2-id,PHENIX-v2-cent,PHOBOS-v2,
  PHOBOS-v2-qm02,STAR-v2-PRC,STAR-v2-PRL},
  characterized
  by the second harmonic moment of the transverse momentum distribution,
  frequently referred to as the ``elliptic flow" and denoted by $v_2$.
  This quantity is determined, for various centrality selections,
  as a function of the transverse mass and particle
  type at mid-rapidity as well as a function of the pseudo-rapidity $\eta
   = 0.5 \log(\frac{|p| + p_z}{|p| - p_z})$.
   Pseudorapidity measures the zenith angle distribution
   in momentum space, but for particles with high momentum,
   $|p|\approx E_{|p|}$, it approximates the rapidity
   $y = 0.5 \log(\frac{E + p_z}{E - p_z})$
   that characterizes the longitudinal momentum distribution
   and transforms additively for longitudinal boosts,
   hence the rapidity density $dn/dy$ is invariant for longitudinal boosts.
   The PHOBOS collaboration found~\cite{PHOBOS-v2},
  that $v_2(\eta)$ is a strongly decreasing function
  of $|\eta|$, which implies
   that the concept of boost-invariance, suggested by
  Bjorken in ref.~\cite{Bjorken} to characterize the physics of
  very high energy heavy ion collisions, cannot be applied to characterize
  the hadronic final state of Au+Au collisions at RHIC.
  A similar conclusion can be drawn from the measurement of the inhomogeneous
  (pseudo)rapidity $dn/d\eta$ and $dn/dy$ distributions of charged particle production at RHIC
  by both the BRAHMS~\cite{BRAHMS-dndeta} and PHOBOS~\cite{PHOBOS-dndeta}
  collaborations, proving the lack of boost-invariance in these reactions, as
  $dn/dy \neq const$ at RHIC.
  Although many models describe successfully
  the transverse momentum dependence of
  the elliptic flow at mid-rapidity, $v_2(p_t, \eta=0)$,
  see ref.~\cite{v2-review} for a recent review on this topic,
  to our best knowledge and an up-to-date
  scanning of the available high energy and nuclear physics literature,
   no model has yet been able to reproduce
   the measured pseudo-rapidity dependence of the elliptic flow at RHIC.

  Hence we present here the first successful attempt to describe
  the pseudo-rapidity dependence of the elliptic flow $v_2(\eta)$ at RHIC. Our tool is the Buda-Lund
  hydrodynamic model~\cite{3d,3d-qm95},
  which we extend here from axial to ellipsoidal symmetry.
  The Buda-Lund hydro model takes into account
  the finite longitudinal extension of the
  particle emitting source, and we show here how the
  finite longitudinal size of the source
  leads naturally to a $v_2$ that decreases with increasing values of
  $|\eta |$, in agreement with the data. We describe simultaneously the
  pseudorapidity and the transverse momentum dependence of the elliptic flow,
  with a parameter set, that reproduces~\cite{ster-ismd03}
  the single-particle transverse momentum
  and pseudo-rapidity distributions as well as the radius parameters of the
  two-particle Bose-Einstein correlation functions, or HBT radii, in case
  when the orientation of the event plane is averaged over. All these benefits
  are achieved with the help of transparent and simple analytic formulas,
  that are natural extensions of our earlier results to the case
  of ellipsoidal symmetry.

\section{Buda-Lund hydro for ellipsoidal expansions}
   The Buda-Lund model is defined with the
  help of its emission function $S(x,p)$,
   where $x = (t, r_x, r_y, r_z)$ is a point in space-time
   and $p = (E, p_x, p_y, p_z)$ stands for the four-momentum.
   To take into account the effects of long-lived resonances,
   we utilize the core-halo model~\cite{chalo}, and characterize
   the system with a hydrodynamically evolving core and a halo of the
   decay products of the long-lived resonances.
   Within the core-halo picture, the measured intercept parameter
   $\lambda_*$ of the two-particle
   Bose-Einstein correlation function is related~\cite{chalo}
   to the strength of the relative contribution
   of the core to the total particle production at a given four-momentum,
\begin{eqnarray}
   S(x,p) & = & S_c(x,p) + S_h(x,p), \qquad \textnormal{and}\\
   S_c(x,p) & = & \sqrt{\lambda_*} S(x,p).
\end{eqnarray}
   Based on the success of the Buda-Lund hydro model to describe
  $Au+Au$ collisions at RHIC~\cite{bl-rhic,ster-ismd03}, $Pb+Pb$ collisions
   at CERN SPS~\cite{bl-sps} and
   $h+p$ reactions at CERN SPS~\cite{bl-na22,cs-rev},
  we assume that the core evolves in a hydrodynamical manner,
\begin{equation}
   S_c(x,p) d^4 x = \frac{g}{(2 \pi)^3}
    \frac{ p^\mu d^4\Sigma_\mu(x)}{B(x,p) +s_q},
\end{equation}
   where $g$ is the degeneracy factor ($g = 1$ for identified pseudoscalar mesons,
   $g = 2$ for identified spin=1/2 baryons),
   and $p^\mu d^4 \Sigma_\mu(x)$ is a generalized Cooper-Frye term,
   describing the flux of particles through
   a distribution of layers of freeze-out hypersurfaces,
   $B(x,p)$ is the (inverse) Boltzmann phase-space distribution,
   and the term $s_q$ is determined by quantum statistics,
   $s_q = 0$, $-1$, and $+1$ for Boltzmann, Bose-Einstein and Fermi-Dirac
   distributions, respectively.

  For a hydrodynamically expanding system, the (inverse) Boltzmann
   phase-space distribution is
\begin{equation}B(x,p)=
  \exp\left( \frac{ p^\nu u_\nu(x)}{T(x)} -\frac{\mu(x)}{T(x)} \right).
\end{equation}
  We will utilize some ansatz for the shape of the flow four-velocity,
  $u_\nu(x)$, chemical potential, $\mu(x)$, and temperature, $T(x)$
  distributions. Their form is determined with
  the help of recently found exact solutions of hydrodynamics,
  both in the relativistic~\cite{relsol-cyl,relsol-ell,relsol-csh}
  and in the non-relativistic cases~\cite{nr-sol,nr-ell,nr-inf},
  with the conditions that these distributions
  are characterized by mean values and variances,
  and that they lead to (simple) analytic formulas
  when evaluating particle spectra and two-particle correlations.

   The generalized Cooper-Frye prefactor is
  determined from the assumption that the freeze-out happens, with
  probability $H(\tau) d\tau$, at
  a hypersurface characterized by $\tau=const$ and
    that the proper-time measures the time elapsed in a fluid element
  that moves together with the fluid, $d\tau = u^\mu(x) dx_\mu$.
  We parameterize this hypersurface with the coordinates $(r_x,r_y,r_z)$
  and find that $d^3 \Sigma^\mu(x|\tau) = u^\mu(x) d^3 x/ u^0(x)$.
  Using $\partial_t \tau|_r = u^0(x)$ we find that in this case
  the generalized Cooper-Frye prefactor is
\begin{equation}
  p^\mu d^4 \Sigma_\mu(x) =p^\mu u_\mu(x) H(\tau) d^4 x,
\end{equation}
  This finding generalizes the
  result of ref.~\cite{cracow} from the case of a spherically
  symmetric Hubble flow to anisotropic, direction dependent Hubble
  flow distributions.

   From the analysis of CERN SPS and RHIC data~\cite{bl-sps,bl-rhic,ster-ismd03},
  we find that the proper-time distribution in heavy ion collisions
  is rather narrow, and $H(\tau)$ can be well
  approximated with a Gaussian representation of the Dirac-delta
  distribution,
\begin{equation}
  H(\tau) = \frac{1}{(2 \pi \Delta\tau^2)^{1/2}}
  \exp\left(-\frac{(\tau - \tau_0)^2 }{ 2 \Delta \tau^2}\right),
\end{equation}
   with $\Delta \tau \ll \tau_0$.

   Based on the success of the Buda-Lund hydro model to describe the axially
   symmetric collisions, we develop an ellipsoidally symmetric extension of the
   Buda-Lund model, that goes back to the successful axially
   symmetric case~\cite{3d,3d-qm95,bl-sps,bl-rhic,ster-ismd03}
   if axial symmetry is restored, corresponding to the
  $X = Y$ and $\dot X = \dot Y$ limit.

   We specify a fully scale invariant,
  relativistic form, which reproduces
   known non-relativistic hydrodynamic solutions too,
   in the limit when the expansion is non-relativistic.
   Both in the relativistic and the non-relativistic cases, the ellipsoidally
   symmetric, self-similarly expanding hydrodynamical solutions can be formulated in
   a simple manner, using a scaling variable $s$ and a corresponding
   four-velocity distribution $u^\mu$, that satisfy
\begin{equation}
  u^\mu \partial_\mu s = 0, \label{e:scalingvar}
\end{equation}
  which means that $s$ is a good scaling variable
   if its co-moving derivative vanishes~\cite{relsol-cyl,relsol-ell}.
   This equation couples the scaling variable $s$ and the flow velocity
   distribution.
   It is convenient to introduce the dimensionless,
   generalized space-time rapidity variables $(\eta_x,\eta_y,\eta_z)$,
   defined by the identification of
\begin{equation}
   (\sh \eta_x, \, \sh \eta_y, \, \sh\eta_z) =
   (r_x \frac{\dot X}{X}, \, r_y \frac{\dot Y}{Y}, \, r_z \frac{\dot Z}{Z}).
\end{equation}
   Here $(X,Y,Z)$ are the characteristic sizes (for example, the lengths of
   the major axis) of the expanding ellipsoid, that depend on proper-time $\tau$ and
   their derivatives with respect to proper-time are denoted by
  $(\dot X, \dot Y, \dot Z)$.
   The distributions will be given in this $\eta_i$ variables, but the
   integral-form is the standard $d^4 x = dt dr_x dr_y dr_z$,
  so we have to take a Jacobi-determinant into account.
   Eq.~(\ref{e:scalingvar}) is satisfied by the choice of
\begin{eqnarray}
  s & = & \frac{\cosh \eta_x -1}{\dot X^2_f}
    +\frac{\cosh \eta_y -1}{\dot Y^2_f}
    +\frac{\cosh \eta_z -1}{\dot Z_f^2}, \\
  u^\mu & = & (\gamma, \, \sinh \eta_x, \, \sinh \eta_y, \, \sinh\eta_z),
\end{eqnarray}
   and from here on
  $(\dot X_f, \dot Y_f, \dot Z_f)
  = (\dot X(\tau_0), \dot Y(\tau_0), \dot Z(\tau_0) )
  = (\dot X_1, \dot X_2,\dot X_3)$,
   assuming that the rate of expansion is constant in the narrow proper-time
   interval of the freeze-out process. The above form has the desired
   non-relativistic limit,
\begin{equation}
  s \rightarrow
    \frac{r_x^2}{2 X_f^2} + \frac{r_y^2}{2 Y_f^2} +\frac{r_z^2}{2 Z_f^2},
\end{equation}
   where again
   $(X_f, Y_f, Z_f) = (X(\tau_0), Y(\tau_0), Z(\tau_0))=(X_1,X_2,X_3)$.
   From now on, we drop subscript $_f$.
   From the normalization condition of $u^\mu(x) u_\mu(x) = 1$ we obtain
   that:
\begin{equation}
  \gamma = \sqrt{ 1 + \sh^2\eta_x+ \sh^2\eta_y+ \sh^2\eta_z},
\end{equation}
   For the fugacity distribution we assume a shape, that leads to
   Gaussian profile in the non-relativistic limit,
\begin{equation}
   \frac{\mu(x)}{T(x)} = \frac{\mu_0}{T_0} - s,
\end{equation}
   corresponding to the solution discussed
  in refs.~\cite{nr-sol,nr-ell,csorgo-ellobs}.
   We assume that the temperature may depend on the position as
   well as on proper-time.
  We characterize the inverse temperature distribution
   similarly to the shape used in the axially symmetric model of
  ref.~\cite{3d,3d-qm95}, and discussed in the exact hydro
  solutions of refs~\cite{nr-sol,nr-ell},
\begin{equation}\frac{1}{T(x)}= \frac{1}{T_0} \left( 1 +
   \frac{T_0 - T_s}{T_s} \:s\right) \left( 1 + \frac{T_0 - T_e}{T_e} \,
   \frac{(\tau -\tau_0)^2}{2 \Delta\tau^2}\right)
\end{equation}
   where $T_0$, $T_s$ and $T_e$ are the temperatures of the center,
   and the surface at the mean freeze-out time $\tau_0$,
   while $T_e$ corresponds to the temperature of the
   center after most of the particle emission is over (cooling due to evaporation
   and expansion). Sudden emission corresponds to $T_e = T_0$, and the
   $\Delta\tau \rightarrow 0$ limit.
  It is convenient to introduce the following quantities:
\begin{equation}
  a^2 = \frac{T_0 - T_s}{T_s} = \left< \frac{\Delta T}{T} \right>_r,
  \label{e:asquare}
\end{equation}
\begin{equation}
  d^2 = \frac{T_0 - T_e}{T_e} = \left< \frac{\Delta T}{T} \right>_t.
\end{equation}

In the above approach we assume the validity of the concept of
local thermalization and the concept of ellipsoidal symmetry at the time of
particle production. We do not know exactly, what freeze-out condition
is realized in Nature. Our above formulas can be also considered as
 a general, second order Taylor expansion of the inverse temperature 
and logarithmic fugacity distributions, while maintaining ellipsoidal symmetry.
We attempt to determine the coefficients of this Taylor expansion from the data in
the subsequent parts. As the saddle-point calculation presented below is sensitive only to
second order Taylor coefficients, any model that has similar second order
expansion leads to similar results. A more theoretical approach is to solve
relativistic hydrodynamics for ellipsoidally expanding fireballs and to
apply the presently most advanced freeze-out criteria, for example
the method of escaping probabilities developed by Akkelin, Hama and Sinyukov
in ref.~\cite{Sinyukov:2002if}.

We also note that the applied distribution of freeze-out temperatures
goes back to the dynamical calculation of ref.~\cite{Csorgo:dd}.
This calculation starts from an initial condition at $\tau = 3.5 $ fm/c, which was 
given by a Parton Cascade Model (PCM) calculation 
for Au+Au collisions at $\sqrt{s_{NN}}=200$ GeV.
This initial condition is 
followed by a three dimensional Hubble flow. In addition, this
dynamical calculation determined the thermodynamical
constraints of entropy and local energy-momentum conservation 
for a sudden time-like deflagration from a supercooled quark-gluon plasma (QGP) phase
to a pion gas phase. Such a transition may start at a constant proper-time,
at some (position independent) value of the local temperature,
that corresponds to the supercooled quark-gluon plasma (QGP) phase.
Such sudden hadronization from a supercooled QGP
phase can be realized due to the large characteristic nucleation times of hadronic
bubbles inside a supercooled QGP, see ref.~\cite{Csorgo:dd} for details.
The strong three-dimensional expansion leads to
negative pressures, mechanical instabilities and the resulting time-like deflagration
may produce hadronic final states in the hatched area of figure 1 of ref.~\cite{Csorgo:dd}.
Half of this area corresponds to superheated hadron gas, with {\it hadronic}
temperatures in the range of 1.0 - 1.4 $T_c$,
 while half of available final states are hadron gas states below the critical temperature,
with hadronic temperatures of 0.8 - 1.0 $T_c$. Thus it is reasonable to assume, that in a
realistic situation, the hadronic final states correspond to a range of temperatures
even if the transition starts from the same temperature everywhere on a constant
proper-time hypersurface. The variation can only be increased if the temperature
of the prehadronic state happened to be inhomogeneous on a constant proper-time hypersurface.
This is why we have assumed in the present paper,
that the hadronic temperatures may be position dependent on a constant proper-time
hypersurface. Instead of using a priori assumptions, we determine 
the parameter values of the hadronic local temperature
distribution a posteriori, from recent data on Au+Au collisions at RHIC.

\section{Integration and saddle-point approximation}
   The observables are calculated analytically from the Buda-Lund
   hydro model, using a saddle-point
   approximation in the integration. This approximation is exact both in the
   Gaussian and the non-relativistic limit,
  and if $p^\nu u_\nu / T \gg 1$ at the point of
   maximal emittivity.
   In this approximation, the emission function looks like:
\begin{equation}
  S(x,k) d^4 x = \frac{g}{(2 \pi)^3}
  \frac{p^\mu u_\mu(x_s)\, H(\tau_s) }{B(x_s,p)+s_q} \,
    \exp\left(-R^{-2}_{\mu \nu} (x-x_s)^\mu(x-x_s)^\nu \right)d^4x
\end{equation}
where
\begin{equation}R^{-2}_{\mu \nu} =
   \partial_{\mu}\partial_{\nu} (- \ln(S_{0}))_s,
\end{equation}
   and $x_\nu$ stands here for $(\tau,r_x,r_y,r_z)$. In the integration,
   a Jacobian ${\tau}/{t}$ has to be introduced when changing the
   integration measure from $d^4x$ to $d\tau d^3x$.

   The position of the saddle-point can be calculated from the equation
\begin{equation}
  \partial_{\mu} (- \ln(S_{0}))(x_s,p) = 0.
\end{equation}
   Here we introduced $S_{0}$, as the 'narrow' part of the emission function:
\begin{equation}S_{0}(x,p)=
   \frac{H(\tau)}{B(x,p)+s_q}.
\end{equation}
   In general, we get the following for the saddle-point
   in$(\tau,\eta_x,\eta_y,\eta_z)$ coordinates:
\begin{eqnarray}
  \tau_s & = & \tau_0, \\
  \sinh \eta_{i,s} & = & \frac{p_i \dot X_i^2
   \cosh\eta_{i,s}}{T(x_s) \left(1+a^2\frac{p_\mu u^\mu(x_s)}{T_0}\right)+p_0
   \frac{\cosh \eta_{i,s}}{\gamma(x_s)}\dot X_i^2}. \label{e:etais}
\end{eqnarray}The
   system of equations~(\ref{e:etais}) can be solved efficiently for the
   saddle-point positions $\eta_{s,i}$ using a successive approximation.
   This method was implemented in our data analysis.
   For the widths of the distributions, we obtained exact analytic results,
   given in terms of $\eta_{s,i}$ that we determined from the successive
   approximations. We have used the full expressions, obtained simply with
  the help of derivations at the saddle-point, for the evaluation of
   the observables and for the comparison with the data. However,
   we summarize here only the leading order approximative result, due to
   reasons of clarity and transparency.

  In the simplest case, where all three $\eta_{i,s}$ are small:
\begin{eqnarray}
  \frac{r_{i,s}}{X_i} & = &
   \frac{\frac{p_i}{T_0} \dot X_i}{ 1+\frac{X_i^2}{R_{T,i}^2}}\textnormal{, for } i=x,y,z, \\
  \frac{1}{R_{i,i}^2} & = & \frac{B(x_s,p)}{B(x_s,p)+s_q}
   \;\left(\frac{1}{X_i^2} + \frac{1}{R_{T,i}^2}\right)\label{e:rgeotherm1},\\
  \frac{1}{R_{0,0}^2} & = & \frac{1}{\Delta \tau_{*}^2} =\; \frac{1}{\Delta \tau^2} +
  \frac{B(x_s,p)}{B(x_s,p)+s} \, \frac{1}{\Delta
  \tau_T^2}\label{e:rgeotherm2}.
\end{eqnarray}
  The above
   eqs.~(\ref{e:rgeotherm1}-\ref{e:rgeotherm2}) imply, that the HBT radii are
   dominated by the smaller of the thermal and the geometrical
  length scales in all
   directions, as found in refs.~\cite{3d, 3d-qm95}. Note that the geometrical
   scales stem from the density distribution,
   governed by the fugacity term $\exp[ \mu(x)/T(x) ]$, while the thermal
   lengths originate from the local thermal momentum distribution 
	$\exp[- p^\mu u_\mu(x)/T(x)]$, and in the above limit they are
\begin{eqnarray}
\frac{1}{\Delta \tau_T^2} & = & \frac{m_t}{T_0}
   \,\frac{d^2}{\tau_0^2}, \\
  \frac{1}{R_{T,i}^2} & = & \frac{m_t}{T_0}
   \,\left(\frac{a^2}{X_i^2}+\frac{\dot
   X_i^2}{X_i^2}\right).
\end{eqnarray}

\section{The invariant momentum distribution}

  The invariant momentum distribution can be calculated as
\begin{equation}\mathrm{N_1}(p)=\int d^4 x S(x,p) =
   \frac{1}{\sqrt{\lambda_*}}\int d^4 x S_c(p,x).
\end{equation}
  The result is a simple expression:
\begin{equation}\mathrm{N_1}(p)= \frac{g}{(2 \pi)^3} \overline{E} \;
   \overline{V}\; \overline{C} \;
   \frac{1}{exp\left(\frac{p^\mu u_\mu(x_s)-\mu(x_s)}{T(x_s)} \right)+s_q}, \label{e:imd_blell}
\end{equation}
    where
\begin{eqnarray}
  \overline{E} & = & p_\mu u^\mu(x_s), \\
  \overline{V} & = & (2 \pi)^{3/2}\, \frac{\Delta\tau_*}{\Delta \tau} \,
    \left[\det{R^2_{ij}}\right]^{1/2},\\
  \overline{C} & = & \frac{1}{\sqrt{\lambda_*}}\frac{\tau_s}{t_s}.
\end{eqnarray}
  Let us investigate the structure of this invariant momentum
   distribution, in particular, the exponent of the spectrum.
   Let us introduce
\begin{equation}
  {b(x_s,p)}= \log B(x_s,p),
\end{equation}
  and evaluate this exponent in the limit,
  where the saddle-point coordinates are all small,
\begin{equation}b(x_s,p)= \frac{p_x^2}{2 \overline{m}_t T_{*,x}} +
   \frac{p_y^2}{2\overline{m}_t T_{*,y}}
   +\frac{p_z^2}{2 \overline{m}_t T_{*,z}}
   +\frac{\overline{m}_t}{T_0}
   - \frac{p_t^2}{2 \overline{m}_t T_0} - \frac{\mu_0}{T_0},
\end{equation}
   where the direction dependent slope parameters are
\begin{eqnarray}
   T_{*,x}&=&T_0+\overline{m}_t \, \dot X^2
       \frac{T_0}{T_0 +\overline{m}_t a^2},\\
   T_{*,y}&=&T_0+\overline{m}_t \, \dot Y^2
     \frac{T_0}{T_0 +\overline{m}_t a^2},\\
   T_{*,z}&=&T_0+\overline{m}_t \, \dot Z^2
       \frac{T_0}{T_0 + \overline{m}_t a^2},
\end{eqnarray}
  and 
\begin{equation}
	\overline{m}_t = m_t \ch(\eta_{z,s}-y) . \label{e:mtbar}
\end{equation} 
   In the limit when we neglect the possibility of a temperature inhomogeneity
   on the freeze-out hypersurface, $a=0$, and using a non-relativistic
  approximation of $\overline{m}_t \approx m$, we recover the
  recent result of ref.~\cite{csorgo-ellobs} for the mass dependence of
  the slope parameters of the single-particle spectra:
\begin{eqnarray}
   T_{*,x}&=&T_0+{m} \, \dot X^2, \\
   T_{*,y}&=&T_0+{m} \, \dot Y^2, \\
   T_{*,z}&=&T_0+{m} \, \dot Z^2.
\end{eqnarray}

\section{The elliptic flow}
  Note, that $b(x_s,p)$ is the only part of the IMD,
  that is explicitly angle dependent, so
\begin{eqnarray}\mathrm{N_1}(p) & \sim & \exp\left(-\frac{p_x^2}{2
   \overline{m}_tT_{*,x}} - \frac{p_y^2}{2 \overline{m}_t T_{*,y}}\right) =
   \nonumber\\& = & \exp\left(-\frac{p_t^2}{2 \overline{m}_t T_{\mathit{eff}}}
   +\left(\frac{p_t^2}{2 \overline{m}_t T_{*,x}} - \frac{p_t^2}{2\overline{m}_t
   T_{*,y}}\right)
   \frac{\cos(2\varphi)}{2}\right),
\end{eqnarray}where
\begin{equation}
  T_{\mathit{eff}}=\frac{1}{2}\left(\frac{1}{T_{*,x}}+\frac{1}{T_{*,y}}\right).\label{e:teff}
\end{equation}
  So, we can easily extract the angular dependencies.
    Let us compute $v_2$ by integrating on the angle:
\begin{equation}v_2=\frac{I_1(w)}{I_0(w)},
\end{equation}where
\begin{equation}w=\frac{p_t^2}{4
   \overline{m}_t} \left(\frac{1}{T_{*,y}}
   -\frac{1}{T_{*,x}}\right). \label{e:ww}
\end{equation}
  Generally, we get from the definition
\begin{equation}N_1=\frac{d^3 n}{dp_z p_t dp_t d\varphi} = \frac{d^2
   n}{2 \pi dp_zp_t dp_t} \left[1+2\sum^{\infty}_{n=1}{v_n \cos(n
   \varphi)}\right]
\end{equation}the following equations:
\begin{eqnarray}v_{2n} & =
   & \frac{I_n(w)}{I_0(w)}, \label{e:v2n} \\
   v_{2n+1} & = & 0\textnormal{.} \label{e:v2n1}
\end{eqnarray}As first and the third flow coefficients vanish in
   this case, a tilt angle $\vartheta$ has to be introduced to get results
  compatible with observations, as discussed in the subsequent parts.

  For large rapidities, $|\eta_s - y|$ becomes also large,
  and
	 $\overline{m}_t = m_t \cosh(\eta_s - y)$ 
	diverges, hence
  $w \rightarrow 0$. Thus we find a natural mechanism
  for the decrease of $v_2$ for increasing values of $|y|$,
  as in this limit, $v_2 \rightarrow I_1(0)/I_0(0) = 0.$

\section{Elliptic flow for tilted ellipsoidal expansion}
  Now, let us compute the elliptic flow for tilted,
   ellipsoidally expanding sources, too,
   because we can get a non-vanishing $v_1$ and $v_3$ only
   this way, in case of $\vartheta \not = 0$, similarly to the
  non-relativistic case discussed in ref.~\cite{csorgo-ellobs}.
  The observables are determined in the center of mass frame of
   the collision (CMS), where the $r_x$ axis points to the direction
  of the impact parameter and the $r_z$ axis points to the direction
   of the beam. In this frame, the ellipsoidally expanding fireball,
  described in the previous sections, may be rotated. So let us
  assume, that we re-label all the $x$ and $p$ coordinates
  in the previous parts
  with the superscript~', e.g. $x \rightarrow x'$ and $p \rightarrow p'$,
  to indicate that these calculations were performed in the system of
  ellipsoidal expansion (SEE), where the principal axis of the expanding
   ellipsoid coincide with the principal axis of SEE.
  In the following, we use the unprimed variables to denote
  quantities defined in the CMS, the frame of observation.

  We assume, that the initial conditions of the hydrodynamic evolution
    correspond to a rotated ellipsoid in CMS~\cite{csorgo-ellobs}.
  The tilt angle $\vartheta$ represents the rotation of the major
  (longitudinal) direction of expansion, $r_z'$, from the direction
  of the beam, $r_z$. Hence the event plane is the $(r_x',r_z') $
  plane, which is the same, as the $(r_x,r_z)$ plane.
  The (zenith) angle between directions $r_z$ and $r_z'$ is
  the tilt angle $\vartheta$, while (azimuthal) angle $\varphi$
  is between the event plane and the direction of the transverse
  momentum $p_t$.

  From the invariant momentum
   distribution, $v_m$ can be calculated as follows:
\begin{equation}v_m = \int_0^{2 \pi} {\frac{\frac{dn}{dp_z
   p_t dp_td\varphi}}{\frac{dn}{dp_z p_t dp_t 2 \pi}} cos(m \varphi)
   d\varphi} \label{e:vmtilt}
\end{equation}
  We have made the coordinate transformation
\begin{eqnarray}
p_x' & = & p_x \cos \vartheta - p_z \sin \vartheta, \\
p_y' & = & p_y, \\
p_z' & = & p_z \cos \vartheta + p_x \sin \vartheta,
\end{eqnarray}
and in addition:
\begin{eqnarray}
p_x & = & p_t \cos \varphi,\\
p_y & = & p_t \sin \varphi,
\end{eqnarray}
     and calculated the transverse momentum and the
  pseudorapidity dependence of $v_2$, for a parameter set
  determined from fitting the axially symmetric version of the
  Buda-Lund hydro model to single particle pseudo-rapidity distribution of
  BRAHMS~\cite{BRAHMS-dndeta} and PHOBOS~\cite{PHOBOS-dndeta},
  the mid-rapidity transverse momentum spectra of identified particles
  as measured by PHENIX~\cite{PHENIX-spectra-id-130,PHENIX-spectra-id-200}  and the two-particle
  Bose-Einstein correlation functions or HBT radii as measured by
  the PHENIX~\cite{PHENIX-HBT} and STAR~\cite{STAR-HBT} collaborations.

   We determined the harmonic moment of eq. ~(\ref{e:vmtilt}) numerically,
  for the case of $m=2$,
  but using the analytic expression of eq.~(\ref{e:imd_blell})
   for the invariant momentum distribution,
  computing the coordinates of the
  saddle point with a successive approximation.
  The successive approximation means a loop here instead
  of solving the non-analytic saddle-point equations.
  We have chosen a loop long enough and have checked
  that an even longer loop will not
  modify the results. This was the same with the width of
   the integration-step.
  We integrated $\mathrm{N_1}(p)$ over $p_t$, as the data were
   taken this way, too. Finally, we were able to describe 
  $v_2(\eta=0,p_t)$ and $v_2(\eta)$
   with the same set of parameters.

  The results are summarized both in Fig. 1 and 2. We find that
  a small asymmetry in the expansion gives a natural description of
    the transverse momentum dependence of $v_2$.
  The parameters are taken from the results
  Buda-Lund hydro model fits to the two-particle Bose-Einstein
  correlation data (HBT radii) and the single particle
  spectra of Au + Au collisions at $\sqrt{s_{NN}} = 130$ GeV,
  ref.~\cite{bl-rhic,ster-ismd03}, where the axially symmetric version of the
  model was utilized. Here we have introduced parameters that
  control the asymmetry of the expansion in the $X$ and $Y$ directions
  such a way that the angular averaged, effective source
  is unchanged. For example, we required that the
  effective temperature, $T_{\mathit{eff}}$ of eq.~(\ref{e:teff}) is
  unchanged. We see on Figs. 1 and 2 that this method was successful
  in reproducing the data on elliptic flow,
  with a small asymmetry between the
  two transverse expansion rates.

The identified particle elliptic flow measurement of PHENIX used a method of
determining the reaction plane from the particles at large rapidities, hence its
results are not significantly affected by non-flow correlations, 
see ref.~\cite{PHENIX-v2-id}.
Fig. 1. illustrates the quality of agreement between our Buda-Lund model
calculation and this PHENIX data set.
  From the parameter values corresponding to Fig. 1, we calculate the value of the
  $v_2(\eta=0)$ and find that this
  value is below the published PHOBOS data point at mid-rapidity by 0.02.
Note, that in order to compute $v_2(\eta)$, one has to integrate $N_1(\eta,p_t,\phi)$ 
over $p_t$ first,
and determine the elliptic flow from the $p_t$ integrated, $\eta$ and 
$\phi$ dependent spectra, as the PHOBOS data were taken this way, too. 
Because of this, there is no simple mathematical connection between 
$v_2(p_t,\eta=0)$, and $v_2(\eta)$ as they do not stem from a common
$v_2(p_t,\eta)$ function.

  The PHOBOS collaboration pointed out the possible existence of
  a non-flow contribution in their $v_2$ data, see ref.~\cite{PHOBOS-v2},
  as they did not utilize the fourth order cumulant method to
  determine $v_2$. 
  We attribute the 0.02 difference between the present Buda-Lund model
	calculation and the PHOBOS data point at mid-rapidity  to such a non-flow
  contribution~\cite{borghini}.
	The magnitude of the non-flow contribution
  has been explicitly studied (but in a different acceptance, at mid-rapidity) 
  by the STAR collaboration. STAR
  found that its value is of the order of 0.01 for mid-rapidity
  minimum bias data in the STAR acceptance, ref.~\cite{STAR-v2-PRC}.

  The good description of the $dn/d\eta$ distribution by the Buda-Lund
  hydro model~\cite{bl-rhic,ster-ismd03} is well reflected
  in the good description of shape of the
  pseudo-rapidity dependence of the elliptic flow.
  Thus the finiteness of the expanding fireball
  in the longitudinal direction and the scaling three dimensional
  expansion is found to be responsible for the experimentally
  observed violations of the boost invariance of both the
  rapidity distribution and that of the collective flow $v_2$.

  Furthermore, the parameter values corresponding to Figs. 1 and 2 indicate  a
  high, $T_0 > T_c = 170 $ MeV central temperature, with a cold surface
  temperature of $T_s \approx 105$ MeV. The success of this description
  suggests that a small fraction of pions may be escaping
  from the fireball from a superheated hadron gas,
  which can be considered as an indication,
  that part of the source of Au + Au collisions at RHIC may be
  a deconfined matter with $T> T_c$.

  Let us determine the size of the volume that is above the
  critical temperature. Within this picture, one can find
  the critical value of $s=s_c$ from the relation that
  $T_0/(1 + a s_c) = T_c$. Using $T_0 = 210$ MeV, $T_c = 170 $ MeV,
    and $a = 1$ we find $s_c = 0.235$.
  The surface of the ellipsoid with $T \ge T_c$ is
  given by

\begin{equation}
  \frac{r_x^2}{X_c^2} +
  \frac{r_y^2}{Y_c^2} +
  \frac{r_z^2}{Z_c^2} = 1.
\end{equation}
  The principal axes of the ``critical" ellipsoid
  are given by
  $X_c = X_f \sqrt{s_c} \simeq~4.2$ fm,
  $Y_c = Y_f \sqrt{s_c} \simeq 5.1$ fm,
  $Z_c = Z_f \sqrt{s_c} \simeq 8.5$ fm,
  hence the volume of the ellipsoid with $T> T_c$ is
  $V_c = \frac{4 \pi}{3} X_c Y_c Z_c \approx 753$ fm$^3$.

  Note, however, that the characteristic average or surface temperature
  of the fireball within this model is $T_s = T_0/(1 + a) \approx 105$
  MeV. So the picture is similar to a snow-ball which has
  a melted core inside.

  Our study shows that this picture is consistent with
  the pseudorapidity and transverse mass dependence of $v_2$
  at RHIC in the soft $p_t < 2$ GeV domain, however, it is
  not yet a direct proof of the existence of a new phase.
  Among others, we have to determine precisely the errors on
  the best fit parameters and to determine the confidence levels
  of the fits, which will be a subject of further research.

	In discussing the significance of the results, it is useful
	to compare it to other calculations to find similarities and 
	differences as well as to map out directions for possible further
	research. For us the key point is not the
	good agreement between the model and the data, but the analytic insight and
	the functional relationships between the model parameters and the observables.
	We checked the model against the data only to demonstrate that we are
	on a good track to understand the rapidity dependence of the elliptic and
	higher order flows, but the fine-tuning of the model parameters is
	a subject of further investigations. 

	Our most important result seems to be
	eqs. (\ref{e:mtbar},\ref{e:ww},\ref{e:v2n},\ref{e:v2n1}),
	that explain analytically why all higher order flows vanish at very forward or 
	backward rapidities. Due to the finite longitudinal size of the source,
	the point of maximal emittivity moves to the
	summit of the expanding ellipsoid as the rapidity is increased to high values.
	Due to the Hubble flow, the transverse momentum distribution has negligible
	transverse flow contributions at this point, the local temperature
	plays the dominant role. However, the local temperature contributes equally
	in both transverse directions, hence all second and higher order flows vanish
	at very forward rapidities. Similar observations hold in the very backward
	direction, due to symmetry reasons.

	When comparing to earlier calculations, we observe that the two key features,
	the finite longitudinal size and the three dimensional Hubble flow were not
	present simultaneously in other works as far as we know. Also, the
	temperature variations, the cooling on the surface were not considered
	by other attempts to understand elliptic flow. For example, Hirano and Tsuda
	considered a three-dimensional numerical solution of relativistic hydrodynamics
	in ref.~\cite{Hirano:2002ds}. Their model is not too far from the considerations
	presented here, they have a finite longitudinal extension of the source and
	a well developed transverse flow at mid-rapidity. Their Fig. 9 indicates
	that they obtained a vanishing elliptic flow at very forward and backward
	rapidities. However, they utilized a Bjorken type of initial condition,
	and the concept of a constant freeze-out temperature.
	As a result, their elliptic flow is too flat, too Bjorken-like near mid-rapidity.
	Hirano studied in ref.~\cite{Hirano:2000eu}
	the effects of short lived resonance decays on the rapidity dependence of the
	elliptic flow at SPS energies. His result is that resonance decays
	yield a negative non-flow contribution, ranging from -0.15 at mid-rapidity to
	about -0.05 at forward and backward rapidities. We did not explore the consequences
	of such an effect here, however, its magnitude is about the size of the error bars
	on the PHOBOS data. This is one of the interesting directions that can be explored
	in further studies, and the importance of this effect will increase at RHIC as more and
	more high statistics, precision measurements will be available on the elliptic flow.

\section{Summary and conclusions}
  We have generalized the Buda-Lund hydro model to the case of
  ellipsoidally symmetric expanding fireballs.
  We kept the parameters determined from fits to the single particle
  spectra and the two-particle Bose-Einstein correlation functions
  (HBT radii)~\cite{bl-rhic,ster-ismd03},
  and interpreted them as angular averages for the
  direction of the reaction plane.
  Then we found that a small splitting between the expansion rates
  parallel and transverse to the direction of the impact parameter,
  as well as a small zenith tilt of the particle emitting source
  is sufficient to describe simultaneously the
  transverse momentum dependence of the collective flow
  of identified particles~\cite{PHENIX-v2-id} 
at RHIC. If a constant
	non-flow parameter of 0.02 is added to the calculated values,
	we also describe well the pseudorapidity dependence of the collective flow
  ~\cite{PHOBOS-v2,PHOBOS-v2-qm02}.

  The results confirm the indication for quark deconfinement at
  RHIC found in refs.~\cite{bl-rhic,ster-ismd03},
  based on the observation, that some of the particles are emitted
  from a region with higher than the critical temperature,
  $T > T_c = 170$ MeV. We estimated that the size of this
  volume is about 1/8-th of the total volume measured
  on the $\tau=\tau_0$ main freeze-out hypersurface,
  totaling of about 753 fm$^3$.
  However, the analysis indicates that the average or
  surface temperature is rather cold, $T_s \approx 105$ MeV,
  so approximately 7/8 of the particles are emitted from a rather cold
  hadron gas.

\section*{Acknowledgments}
  T. Cs. would like to thank B. L\"orstad and G. Gustafson
   for kind hospitality
  during his stay at the University of Lund in spring 2003.
  The support of the following grants are gratefully acknowledged:
  OTKA T038406, T034269, the exchange programs  of the
  OTKA, MTA and NSF under grant INT0089462 and the Hungarian and Polish Academy of
  Sciences, as well as the NATO PST.CLG grant 980086.

\begin{figure}
\begin{center}
\includegraphics*[scale = 0.9]{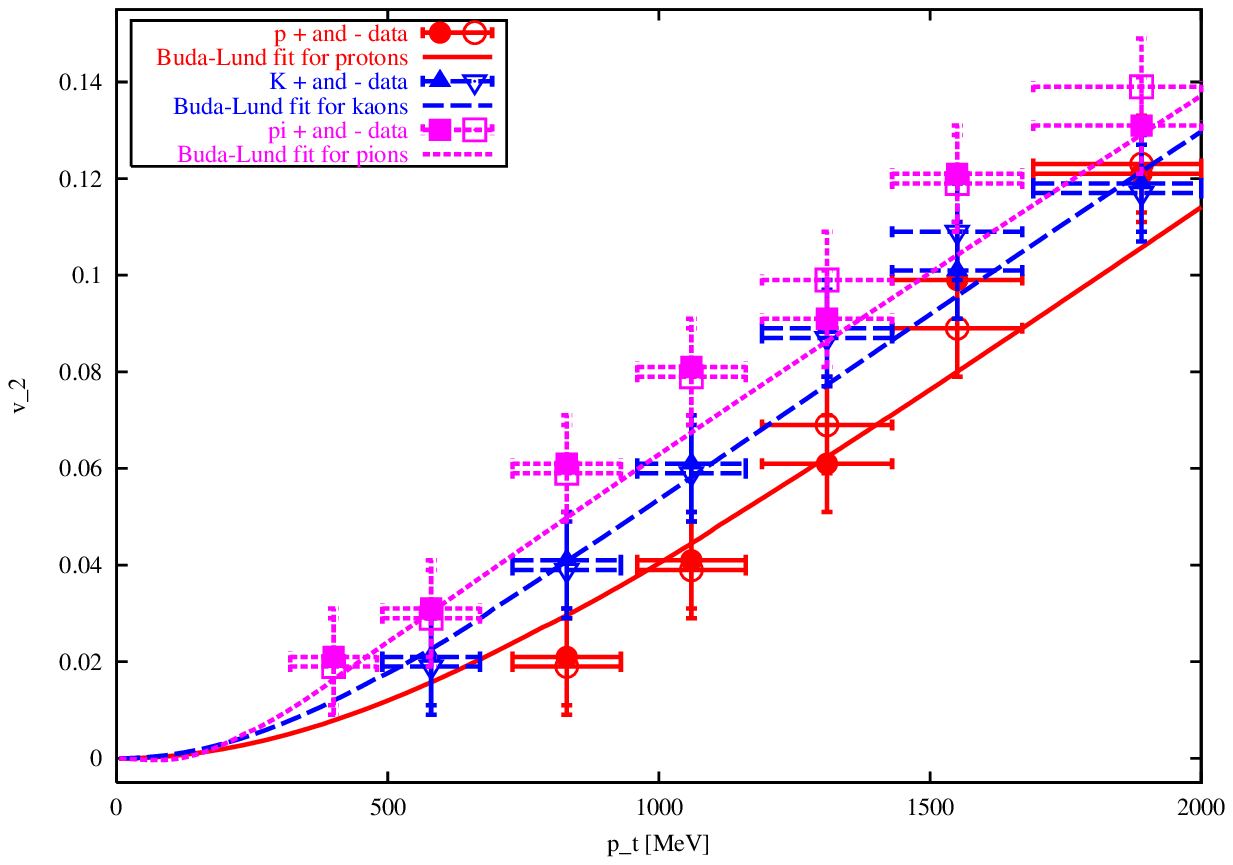}
\end{center}
\caption{Buda-Lund model calculation and the $v_2(p_t,\eta=0)$ data}\label{v2pt}
\small{Ellipsoidally symmetric Buda-Lund calculation compared to the PHENIX $v_2(p_t)$ data of
identified particles~\cite{PHENIX-v2-id}. The parameter set is:
$T_0=210\textrm{ MeV}$, $\dot X=0.57$, $\dot Y=0.45$, $\dot
Z=2.4$, $a=1$, $\tau_0 = 7$ fm/c, $\vartheta=0.09$, $X_f=8.6$ fm,
$Y_f=10.5$ fm, $Z_f=17.5$ fm, $\mu_{0,\pi}=70\textrm{ MeV}$,
$\mu_{0,K}=210 \textrm{ MeV}$ and $\mu_{0,p}=315 \textrm{ MeV}$,
and the masses are taken as their physical value.}
\end{figure}

\begin{figure}
\begin{center}
\includegraphics*[scale = 0.9]{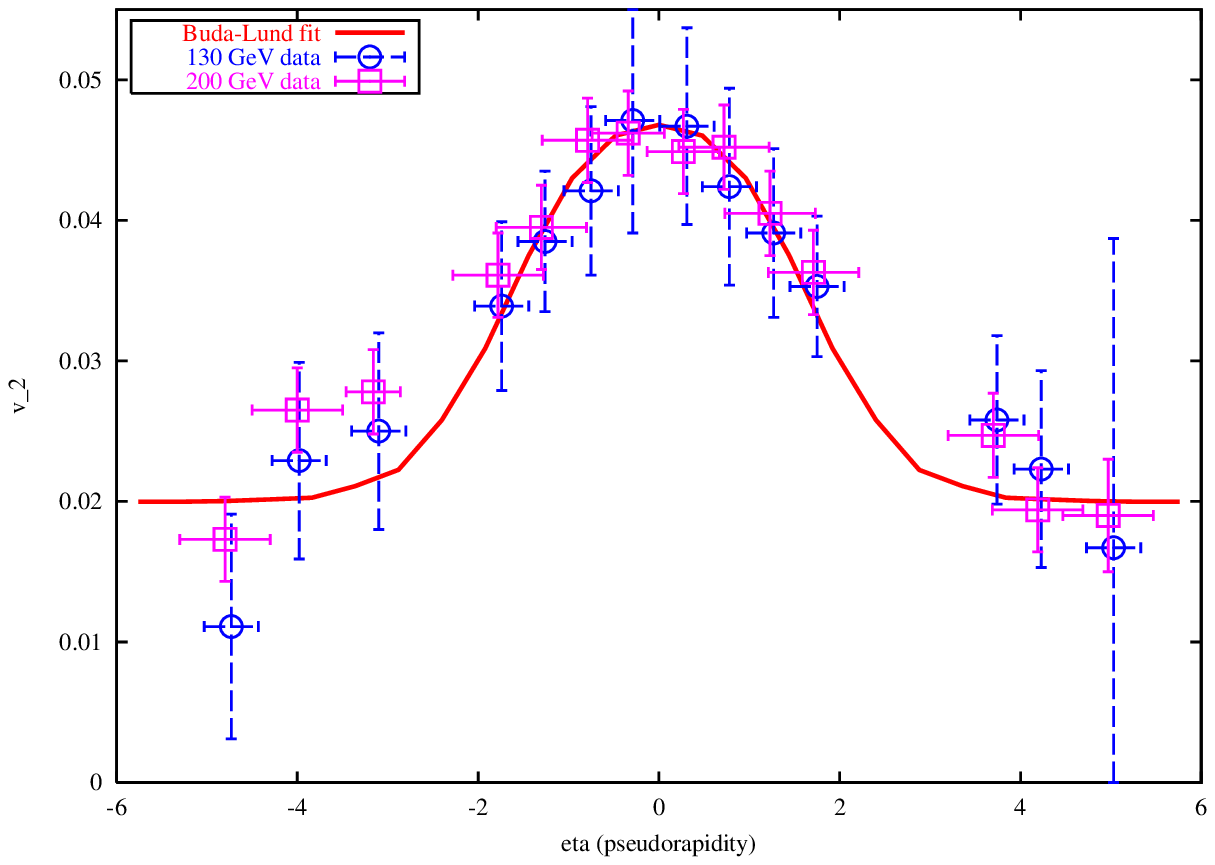}
\end{center}
\caption{Buda-Lund model calculation and the $v_2(\eta)$ data}\label{v2y}
\small{This image compares the ellipsoidally symmetric Buda-Lund model to the $130 \textrm{ GeV}$
   Au+Au and $200 \textrm{ GeV}$ Au+Au $v_2(\eta)$
  data of PHOBOS~\cite{PHOBOS-v2,PHOBOS-v2-qm02}.
Here we used the same parameter set as at fig. \ref{v2pt}, with pion mass and chemical potential, 
and added a  constant non-flow parameter of 0.02 to the calculated values of $v_2$.}
\end{figure}

\vfill\eject

\begin{table}
\begin{center}
\begin{tabular}{|c|l|}
\hline
Parameter & Value \\ \hline \hline
$T_0$ & 210 MeV\\
$\dot X$ & 0.57\\
$\dot Y$ & 0.45\\
$\dot Z$ & 2.4\\
$a$ & 1\\
$\tau_0$ & 7 fm/c\\
$\Delta \tau$ & 0 fm/c\\
$\vartheta$ & 0.09\\
$X_f$ & 8.6 fm\\
$Y_f$ & 10.5 fm\\
$Z_f$ & 17.5 fm\\
$\mu_{0,\pi}$ & 70 MeV\\
$\mu_{0,K}$ & 210 MeV \\
$\mu_{0,p}$ & 315 MeV \\ \hline
\end{tabular}
\end{center}
\caption
{
The table shows the parameter set used to describe the data.
Note, that $a^2 = \frac{T_0 - T_s}{T_s}$,
see equation~(\ref{e:asquare}).
}
\label{tab:results}
\end{table}

\vfill\eject

\end{document}